\documentclass{PoS}

\title{Local gauge symmetry on optical lattices?}

\ShortTitle{Local gauge symmetry on optical lattices?}

\author{ \speaker{Yuzhi Liu}$^{a,b}$\footnote{email: yuzhi-liu@uiowa.edu} , 
        Yannick Meurice$^a$\footnote{email: yannick-meurice@uiowa.edu} , 
        Shan-Wen Tsai $^c$\footnote{email: shan-wen.tsai@ucr.edu}\\
 \llap{$^a$} Department of Physics and Astronomy, University of Iowa, Iowa City, IA 52240, USA\\
 \llap{$^b$} Fermi National Accelerator Laboratory, Batavia, IL 60510, USA \\
  \llap{$^c$} Department of Physics and Astronomy, University of California, Riverside, CA 92521, USA\\}

\abstract{The versatile technology of cold atoms confined in optical lattices allows the creation of a vast number of lattice geometries and interactions, 
providing a promising platform for emulating various lattice models. This opens the possibility of letting nature take care of sign problems and 
real time evolution in carefully prepared situations. Up to now, experimentalists have succeeded to implement several types of Hubbard models considered 
by condensed matter theorists. In this proceeding, we discuss the possibility of extending this effort to lattice gauge theory. We report recent efforts to 
establish the strong coupling equivalence between the Fermi Hubbard model and SU(2) pure gauge theory in 2+1 dimensions by standard determinantal methods 
developed by Robert Sugar and collaborators. We discuss the possibility of using dipolar molecules and external fields to build models where the equivalence 
holds beyond the leading order in the strong coupling expansion.}

\FullConference{The 30th International Symposium on Lattice Field Theory\\
                 June 24 - 29,  2012\\
                 Cairns, Australia}

\begin{document}

\section{Introduction}

New experimental techniques to confine cold atoms in optical lattices have allowed the creation of many lattice geometries and interactions. Trapping polarizable atoms or molecules in a periodic potential created by crossed counterpropagating laser beams has been an area of intense activity in recent years. It is now possible to physically build lattice systems where the average number of particles per site and their tunneling between neighbor sites can be adjusted experimentally. 
 This has opened the possibility of engineering experimental setups that mimic lattice Hamiltonians used by theorists and to follow their real time evolution.
Up to now, experimentalists have succeeded to implement several types of Hubbard models.  Can these successes be extended to lattice gauge theory?

There are problems that are very difficult to address by conventional (digital, classical) computational methods but could be dealt with if lattice gauge models 
could be studied from approximate implementations on optical lattices: for instance 
the phase diagram of QCD at finite temperature and significant chemical potential, real time evolution and string breaking. 
Beyond its success in particle physics, quantum gauge theory also have played an important role in the development of quantum computing. 
Decoherence can be an important obstacle for quantum computation. One of the first applications of Shor's idea of  fault-tolerant quantum computation \cite{Shor:1996qc} was 
Kitaev's toric code \cite{Kitaev:1997wr} where the two stabilizers (the plaquette and the cross) are operators that were introduced two decades before in the Hamiltonian formulation of lattice gauge theory \cite{Fradkin:1978th}. For these reasons, it seems that some effort 
should be put in finding optical lattice realizations of quantum gauge theories. 

It has been suggested \cite{ylm} that the approximate emergent local $SU(2)$ symmetry \cite{anderson,PhysRevB.38.2926} of the Hubbard model at strong coupling could be a good starting point for this enterprise. This is briefly reviewed in Sec. \ref{sec:app}. 
However, the correspondence Hubbard $\simeq$ Heisenberg $\simeq$ Hamiltonian Lattice Gauge Theory only works up to {\it  second order} in the strong coupling expansion and there is no obvious plaquette interactions. The fermions coupled to the gauge fields have color indices but no Dirac spin indices.
The strong coupling expansion could be used as a guide to add interactions or combine different types of atoms or molecules in order to build Hubbard-like model corresponding more accurately to existing lattice gauge models. Some elements of the expanding toolbox that can be used for this purpose are provided in Sec. \ref{sec:toolbox}. Finally, one should 
use computational methods developed in lattice gauge theory to calculate determinants that have been adapted for Hubbard models \cite{Scalettar:1987zz} 
to test the validity of the approach. 

\section{Approximate local symmetry, mesons and baryons in Hubbard models }
\label{sec:app}

The Hubbard model Hamiltonian is 

\begin{equation}
H = -t \sum_{\langle i,j \rangle,\alpha}( c^{\dagger}_{i,\alpha} c^{}_{j,\alpha}+ h.c.) + U \sum_{i=1}^{N} n_{i\uparrow} n_{i\downarrow}
\end{equation}
where $t$ characterizes the tunneling between nearest neighboor sites and $U$ controls the onsite Coulomb repulsion. 

\begin{figure}[h]
\begin{center}
\includegraphics[width=0.7\textwidth,keepaspectratio=]{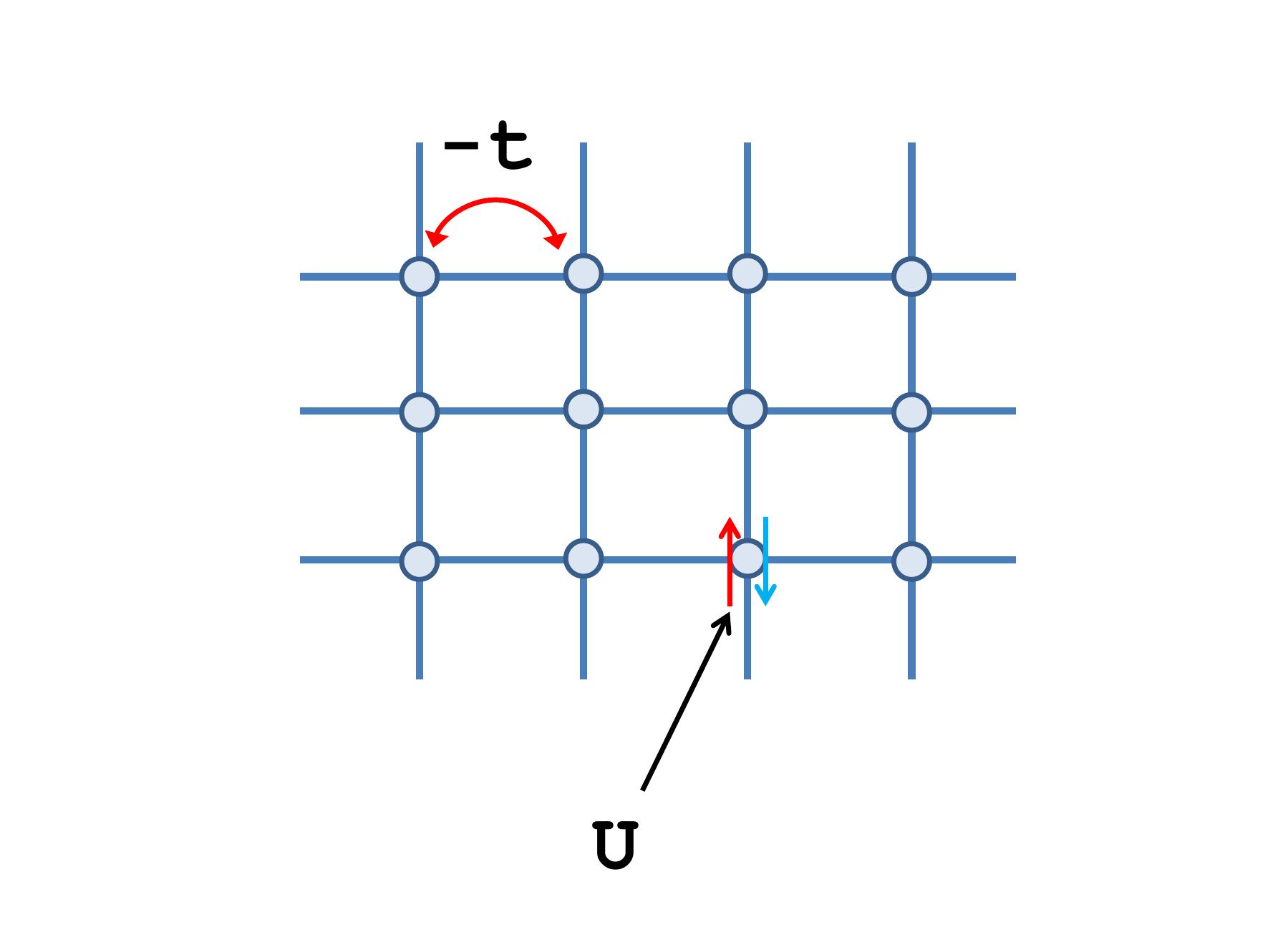}
\end{center}
\caption{Graphical representation of the interations of the Hubbard model. }
\end{figure}

In the strong coupling limit ($U \gg t$) and at half-filling, which has one fermion per site, the previous model is equivalent to Heisenberg model
\begin{equation}
H = J\sum_{<\mathbf{i}\mathbf{j}>} \mathbf{S}_\mathbf{i}\cdot \mathbf{S}_\mathbf{j}
\end{equation}
with $J=4t^2/U$.
The spin operator $\mathbf{S}_\mathbf{i}$ can be written as 
\begin{equation}
\mathbf{S}_\mathbf{i} = \frac{1}{2} f_{\mathbf{i}\alpha}^\dagger \mathbf{\sigma}_{\alpha \beta} f_{\mathbf{i}\beta}
\end{equation}
The corresponding Heisenberg Hamiltonian becomes
\begin{equation}
H=\sum_{<\mathbf{i}\mathbf{j}>}-\frac{1}{2} J f_{\mathbf{i}\alpha}^\dagger f_{\mathbf{j}\alpha} f_{\mathbf{j}\beta}^\dagger f_{\mathbf{i}\beta} +\sum_{<\mathbf{i}\mathbf{j}>} J (\frac{1}{2} n_\mathbf{i}-\frac{1}{4} n_\mathbf{i} n_\mathbf{j})
\end{equation}
A constraint must be imposed in order to recover the original Heisenberg model: $f_{\mathbf{i}\alpha}^\dagger f_{\mathbf{i}\alpha} =1$.

\begin{figure}[h]
\begin{center}
\includegraphics[width=0.7\textwidth,keepaspectratio=]{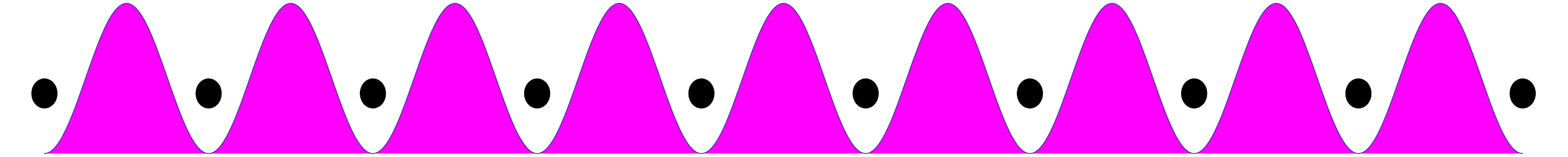}
\end{center}
\caption{Graphical representation of the half-filling, strong onsite repulsion situation.}
\end{figure}

After a particle-hole transformation in the spin down operator \cite{anderson,PhysRevB.38.2926,Tan},
\begin{equation}
f_{\mathbf{i},\uparrow},f^\dagger_{\mathbf{i},\uparrow} \rightarrow \Psi_{x,1},\Psi^\dagger_{x,1};\hspace{0.5cm}f_{\mathbf{i},\downarrow},f^\dagger_{\mathbf{i},\downarrow} \rightarrow \Psi^\dagger_{x,2},\Psi_{x,2}
\end{equation}
The Heisenberg Hamiltonian can be written exactly as follows
\begin{equation}
H= \frac{J}{8} \sum_{x,\hat{\mathbf{i}}}[M_x M_{x+\hat{\mathbf{i}}}+2(B_x^\dagger B_{x+\hat{\mathbf{i}}}+B_{x+\hat{\mathbf{i}}}^\dagger B_{x})]-\frac{Jd}{4}\sum_x(M_x-\frac{1}{2}) \ ,
\label{eq:29}
\end{equation} 
where the ``meson'' and ``baryon'' operators are
\begin{equation}
M_x = \sum_{a=1,2}\Psi_{x,a}^\dagger \Psi_{x,a}
\end{equation}
and
\begin{equation}
B_x = \sum_{a=1,2} \frac{\epsilon_{ab}}{2} \Psi_{x,a} \Psi_{x,a} = \Psi_{x,1} \Psi_{x,2} 
\end{equation}

The anti-ferromagnetic Heisenberg model can 
also be obtained from the strong coupling limit of the Kogut-Susskind Hamiltonian of $SU(2)$ lattice gauge theory \cite{PhysRevB.38.2926} which is defined as

\begin{equation}
H=\frac{8}{3J}\sum_{a,\mathbf{x},\mathbf{i}} E^{a}_{\mathbf{x}\mathbf{i}}E^{a}_{\mathbf{x}\mathbf{i}} +\frac{i}{2}\sum_{a,b,\mathbf{x},\mathbf{i}}(\psi^\dagger_{\mathbf{x},a}\hat{U}^{ab}_{{\mathbf{x},\mathbf{i}}}\psi_{\mathbf{x+i},b} -h.c.)
\label{eq:KS}
\end{equation}
with $J=16/3g^2\propto \beta$.

\vspace{0.5in}
The approximate gauge invariance allows to show the equivalence between two apparently very different mean field theory of the Hubbard models (d-wave super conductivity and ``flux phase'') \cite{anderson}. 
The above approximate equivalence has been proposed \cite{ylm} as a starting point to implement lattice gauge theory on optical lattice. However,
the equivalence with H in Eq. (\ref{eq:KS}) is only valid under second order perturbation theory.
There are no plaquette interactions for the above lattice gauge Hamiltonian and there are no spin indices for the gauge fields. Staggered fermion Hamiltonian can be realized by doubling the lattice spacing \cite{Tan}.
Chemical potential generally introduces sign problem \cite{Loh:1990zz} which makes physical simulations desirable. 
\section{A toolbox to engineer plaquette interactions}
\label{sec:toolbox}
In the following, we discuss some tools that can be used to build new interactions in optical lattice systems. 
\subsection{Dipolar fermions in external fields}
\begin{figure}[htp]
\begin{center}
\includegraphics[width=0.7\textwidth,keepaspectratio=]{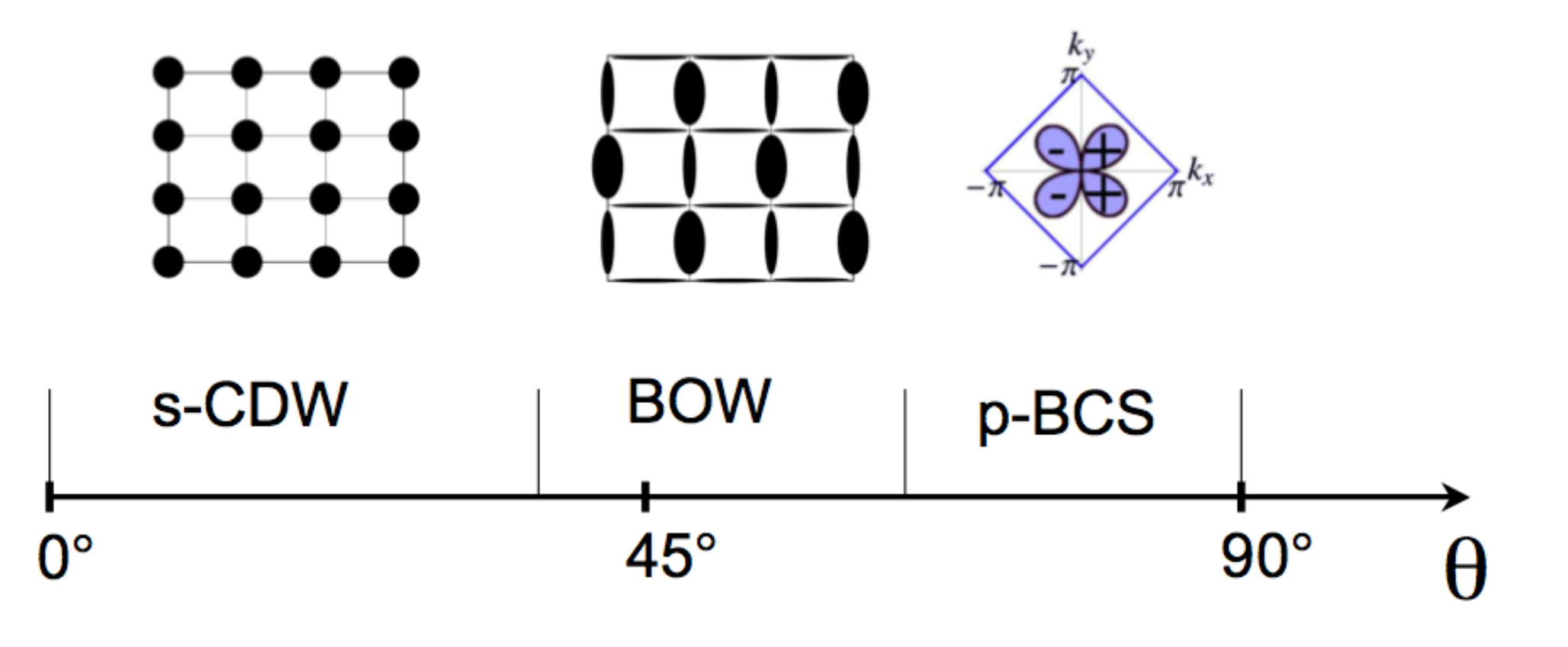}
\end{center}
\caption{\label{fig:fig3}Phase diagram for spinless dipolar fermions on a square lattice at half-filling as a function of the polar angle $\theta$ for the direction of the dipole moments ($\phi = 0$). The CDW phase corresponds to checkerboard order, the BOW has different probabilities on the links, and the $p$-wave BCS has the symmetry of the order parameter as shown on the upper-right corner. }
\end{figure}

External fields can be used to get close to Feshbach resonances or induce dipole interactions 
across sites. From a theoretical point of view, new couplings and long-range interactions can be varied to obtain rich phase diagrams \cite{fradkin}. The orientation of the dipole moments can also be manipulated to tune the dipole interaction. Considering single species fermions on a two-dimensional half-filled square lattice in the xy-plane and fixing the azimuthal angle $\phi = 0$, the variation of the polar angle $\theta$ has been found \cite{shanwen1} to lead to three different phases, as shown in Figure \ref{fig:fig3}. 
For $\theta < \theta_1$, the repulsive nearest-neighbor interaction leads to a conventional density-wave phase, here denoted as CDW (charge-density-wave) 
in analogy to condensed matter systems where the fermions are charged. With increasing $\theta$ ($\theta_1 < \theta < \theta_2$), 
nearest-neighbor interactions along the x-direction become attractive and a density-wave with high-probabilities along the links sets in.
 This phase has been referred to as a bond-order solid (BOS), or a $p$-wave CDW. For $\theta > \theta_2$, the ground state has $p$-wave BCS pairing. Generally, with long-range interactions, ordered states with non-zero angular momentum becomes favored, possibly involving bond orders.
  For spin-$1/2$ fermions, a $p$-wave spin-density wave phase has been found for dipolar fermions on a half-filled square lattice \cite{shanwen2}, corresponding to a vector order parameter on the bonds.

\subsection{Combination of different types of atoms or molecules}
An interesting possibility \cite{andreas} to generate new interactions is to create a bi-partite lattice with S-orbits and P-orbits arranged on a checkerboard pattern and  tune 
the relative heights in order to have the different orbitals at approximately the 
same level. This situation can be used to generate  ring currents which share some feature of the plaquette interactions \cite{cristiane,PhysRevA.82.013616}. 
Another possibility is to use two lattices: one lattice having atoms or molecules that can hop and induce the new 
interactions on the other lattice. 
Cheng Chin's  group uses $^{6}Li$ atoms coupled to $^{133}Cs$. Using a Fermi-Bose mixture of ultra-cold atoms in a optical lattice, one can couple bosonic gauge fields to fermionic matter \cite{zoller}. Models have been constructed by combining quantum link models with spinless staggered fermions \cite{Chandrasekharan:1996ih}. U(1) lattice gauge models could be implemented by using one fermion species and two boson species which represent gauge fields. 

\subsection{Effective plaquette interactions}

It is clear that higher order terms in the strong coupling expansion will generate plaquette interactions. 
More generally, these interactions should appear generically after blocking the Hamiltonian of Eq. \ref{eq:29}.
Recently, an old preprint \cite{orland} explaining how
non-Abelian gauge interactions originate as effective dynamics in models of hopping particles
has been made available on arXiv. 
\section{Work in progress}

We have been working in several directions. 
We started up with the 2 dimensional Hubbard model and 2 + 1
dimensional lattice gauge model with 2 colors. Determinantal calculations to check the approximate equivalence between the two models are in progress. We are writing a Hubbard model code by using the Hybrid Monte Carlo (HMC) algorithm. We plan to add different boson species to the existing code in order to compare with models of Fermion-Boson mixtures confined in the optical lattices.
Progress also needs to be made in higher orders of the strong coupling expansion from Eq. \ref{eq:29} to see how effective plaquette terms can emerge. 
We also need to understand the correspondence between the physical temperature used in real experiments and the coupling ($\beta$ or $g$) used in the lattice gauge model with specific 2+1 geometries.

This work was stimulated by the workshop "Critical Behavior of Lattice Models" in summer 2010 at the Aspen Center for Physics. 
This work was funded in part by the Mathematical and Physical Sciences Funding Program of the University of Iowa.
SWT acknowledges support from NSF under grant DMR- 0847801 and from the UC-Lab FRP under award number 09-LR-05-118602.

\end{document}